\documentclass[sigconf,10pt,screen]{acmart}
\AtBeginDocument{%
  \providecommand\BibTeX{{%
    \normalfont B\kern-0.5em{\scshape i\kern-0.25em b}\kern-0.8em\TeX}}}

\setcopyright{acmcopyright}
\copyrightyear{2023}
\acmYear{2023}
\acmDOI{XXXXXXX.XXXXXXX}

\acmConference[arXiv'23]{}{July 17, 2023}{Online}
\acmPrice{15.00}
\acmISBN{978-1-4503-XXXX-X/18/06}

\makeatletter
\renewcommand{\@copyrightowner}{Copyright held by the owner/author(s). 
This is the author's version of the work.
It is posted here for your personal use. Not for redistribution. 
}
\makeatother

\usepackage{xspace}
\usepackage{graphicx}
\usepackage{booktabs}
\usepackage{multirow}
\usepackage{tabularx}
\usepackage{array}
\usepackage{enumitem}
\graphicspath{{./Assets}}

\usepackage{xcolor}
\usepackage{breakurl}

\usepackage[textfont={normal,bf, it},labelfont=bf]{caption} 
\usepackage[labelformat = parens, labelsep = space, textfont=small]{subcaption}

\definecolor{pro_green}{rgb}{0.0, 0.66, 0.47}

\newcommand{\sys}{\texttt{ExpAR}\xspace}

\newcommand{\1}{{\em (i)}}
\newcommand{\2}{{\em (ii)}}
\newcommand{\3}{{\em (iii)}}
\newcommand{\4}{{\em (iv)}}

\newcommand{\para }[1]{\noindent  {\bf #1}}

\begin{document}

\sloppy

\title{Toward Scalable and Controllable AR Experimentation}

\author{Ashkan Ganj}
\orcid{0009-0006-3490-0471}
\affiliation{%
  \institution{Worcester Polytechnic Institute}
    \streetaddress{}
  \city{}
  \country{}}
\email{aganj@wpi.edu}

\author{Yiqin Zhao}
\orcid{0000-0003-1044-4732}
\affiliation{%
 \institution{Worcester Polytechnic Institute}
 \streetaddress{}
 \city{}
 \state{}
 \country{}
 }
\email{yzhao11@wpi.edu}

\author{Federico Galbiati}
\affiliation{%
 \institution{Worcester Polytechnic Institute}
 \streetaddress{}
 \city{}
 \state{}
 \country{}
 }
\email{fgalbiati@wpi.edu}

\author{Tian Guo}
\orcid{0000-0003-0060-2266}
\affiliation{%
  \institution{Worcester Polytechnic Institute}
  \city{}
  \country{}}
\email{tian@wpi.edu}

\renewcommand{\shortauthors}{Ganj et al.}

\begin{abstract}
To understand how well a proposed augmented reality (AR) solution works, existing papers often conducted tailored and isolated evaluations for specific AR tasks, e.g., depth or lighting estimation, and compared them to easy-to-setup baselines, either using datasets or resorting to time-consuming data capturing. Conceptually simple, it can be extremely difficult to evaluate an AR system fairly and in scale to understand its real-world performance. The difficulties arise for three key reasons: lack of control of the physical environment, the time-consuming data capturing, and the difficulties to reproduce baseline results.

This paper presents our design of an AR experimentation platform, \sys, aiming to provide scalable and controllable AR experimentation. \sys is envisioned to operate as a standalone deployment or a federated platform; in the latter case, AR researchers can contribute physical resources, including scene setup and capturing devices, and allow others to time share these resources. 
Our design centers around the generic sensing-understanding-rendering pipeline and is driven by the evaluation limitations observed in recent AR systems papers. 
We demonstrate the feasibility of this vision with a preliminary prototype and our preliminary evaluations suggest the importance of further investigating different device capabilities to stream in 30 FPS.

The \sys project site can be found at \href{https://cake.wpi.edu/expar}{https://cake.wpi.edu/expar}. 
\end{abstract}



\maketitle

\section{Introduction}

Augmented reality (AR) has emerged as a promising way for users to interact with physical worlds through virtual overlay. For example, in an AR-powered shopping app\footnote{https://www.warbyparker.com/app}, a user can leverage a handheld device and its camera(s) to overlay products of interest, e.g., virtual glasses, on a desirable physical position, e.g., on the user's face~\cite{zhao2023hotmobile}.  
As AR enters the general consumer market, hundreds of millions of users can benefit from this rich media experience with applications ranging from tourism to advertisement~\cite{green_planet_ar,googlexr}.

Over the past decade, we have witnessed a blossom of works that provide high-quality AR performance~\cite{Prakash2019-gb,Yi2020-na,Liu2020-fy,Apicharttrisorn2022Sensys,Guan2022-gv,Zhang2022-lq}, including our work~\cite{Zhao2020-gr-eccv, Zhao2021-mg,Zhao2022-yx}.
However, we have found that it is challenging to fairly and scalably evaluate the developed algorithms and systems to understand their real-world performance. We attribute the evaluation challenge to three key aspects: lack of control of the physical environment, time-consuming data capturing, and difficulty reproducing baseline results. 

This paper introduces \sys, a platform providing scalable and controllable AR experimentation, by centering the key design insight of \emph{generalizable AR pipelines}. Existing practices in capturing experimental data often involve a user interacting with the AR device~\cite{Huzaifa2021-jj,Zhang2022-gg}. However, replicating the user mobility patterns, e.g., walking trajectory and device pose, and the environment information, e.g., scene lighting changes, is difficult. 
\sys controls the physical environment via programmable and remotely controllable mobility sensors, e.g., smart light bulbs and robotic cars~\cite{Liu2023ICRA}. 

Moreover, \sys provides mechanisms to capture high-quality data in a scalable and time-efficient manner. 
Access to high-quality data is crucial for properly evaluating an AR pipeline, from data captured by various sensors (most notably the camera sensors) to render virtual overlay.
For example, prior work demonstrated that a blurred image could impact the accuracy of downstream vision tasks such as image classification and lighting estimation~\cite{Liu2020-fy,Zhao2022-yx}.
\sys can capture high-quality data by supporting various capable hardware devices and carefully controlling their motions.
Additionally, collecting a large amount of data in a scalable and time-efficient way is desirable, though such a desire is not unique to the AR community. For example, the robotic community has recently explored ways to enable multiple remote users to interact with robotic arms~\cite{Tung2021-uj} and collect such data. \sys supports similar remote manipulation and allows programmable and parallel access to the physical capturing devices.  

Lastly, to facilitate reproducibility, \sys breaks the AR pipeline into the sensing, understanding, and rendering steps. For each step, \sys will provide baseline methods that AR researchers can turn on and off to compose the desired AR pipeline dynamically. In other words, \sys will provide the ability to evaluate an AR solution \emph{holistically} in the context of other AR tasks. To improve the baseline diversity, \sys will allow users to upload their own, similar to how Hugging Face hosts DL models~\cite{huggingface}.

\begin{figure}[t]
    \centering
    \includegraphics[width = 0.45 \textwidth]{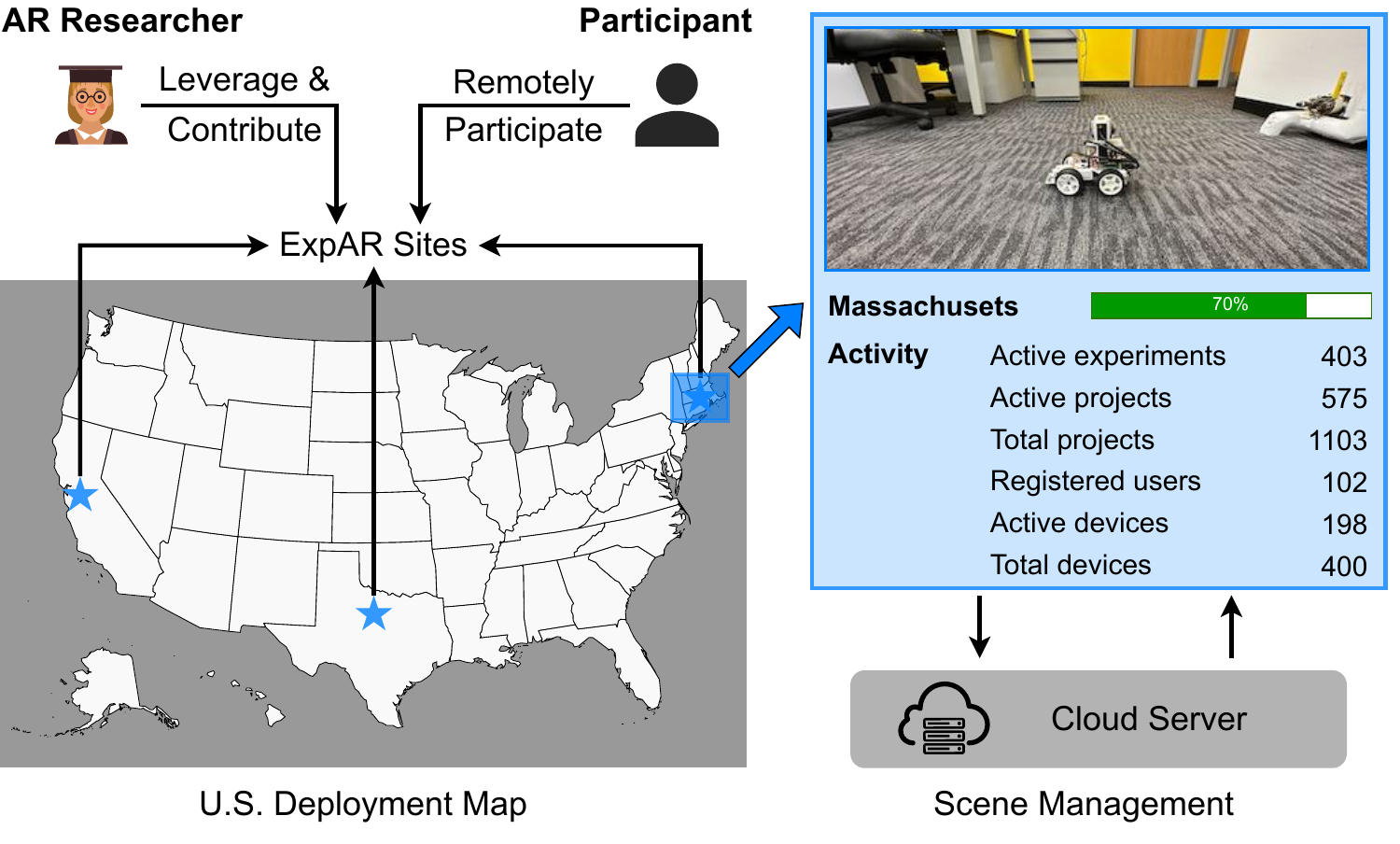}
    \vspace{-.5em}
    \caption{An overview of \sys deployment.
    }
    \vspace{-5mm}
    \label{fig:high_level_overview}
\end{figure}

Figure~\ref{fig:high_level_overview} depicts the high-level overview of \sys,  a fully controllable and programmable AR evaluation platform.
\sys is envisioned to operate as a standalone deployment (which we will describe an initial prototype in \S\ref{sec:impl}) or a federated platform that consists of network-connected deployments at different physical locations, similar to platforms such as PlanetLab and CloudLab~\cite{PlanetLab,cloudlab}. We envision \sys to consist of geographically-dispersed sites. Each site is a physical deployment that includes physical scene setups, AR, and capturing devices, that connect to \sys's backend for data storage and processing.
Both AR researchers and users can interact with the physical setups remotely to carry out key tasks, including scalable data capturing, experiment design, online surveys, and participant observation. 
We make the following key contributions. 
\begin{itemize}[leftmargin=.12in,topsep=0pt]
    \item We pinpoint the limitations of existing AR evaluation methodology via characterizing recent papers and reflecting our evaluation practices.
    \item We describe the design of a fully controllable and programmable AR platform, centering the key insight of decomposable sensing-understanding-rendering AR pipelines. Our design serves as a conceptual framework for implementing a AR researcher-center evaluation platform. 
    \item We present a preliminary prototype and evaluation that showcases the feasibility of programmable visual data capturing, streaming, and storage via a custom-built mobile capturing device and a cloud backend.
\end{itemize}

Our work shares similar spirits with three recent efforts, ILLIXR, XRBench, and CoMIC~\cite{Huzaifa2021-jj,Kwon2022-vm-mlsys,Han2022-bh}, in enabling better support for evaluating the emerging mixed reality applications. 
XRBench focuses on evaluating deep learning models for XR applications in representative execution patterns, which is similar to the design of \sys's \emph{understanding} component.  
While ILLIXR and its multi-user counterpart CoMIC can capture data to evaluate a standalone system, it is not designed with a controllable physical environment, scalable experimentation, and cross-system evaluations in mind. 
In contrast, \sys is designed from outside to address the practical limitations exhibited in evaluating AR systems. 
In short, \sys compliments existing efforts and bridges the gap in reproducible AR research.

\vspace{-3mm}
\section{Limitations of Existing AR Evaluation Methodology}
\label{sec:motivation}

\begin{table*}[!h]
\centering
\caption{A survey of recent AR systems work and their evaluation methodology.
\textnormal{For the last three columns, information inside the parenthesis represents the numerical scale. For example, Y(30) in the user study column means 30 participants.}
}
\label{tab:AR_sys_papers_eval}
\resizebox{0.9\textwidth}{!}{%
\begin{tabular}{@{}l|lccccccc@{}}
\toprule
                       Category   & Paper & Simulation & Task(s) & \multicolumn{1}{c}{\begin{tabular}[c]{@{}c@{}}Visual data capturing\\ (Y*- specialized)\end{tabular}} & \multicolumn{1}{c}{\begin{tabular}[c]{@{}c@{}}Variation\\ (\textbf{S}patial, \textbf{T}emporal)\end{tabular}} & \multicolumn{1}{c}{\begin{tabular}[c]{@{}c@{}}Scene Diversity\\ (\textbf{L}ow, \textbf{M}edium)\end{tabular}} & User Study & \multicolumn{1}{c}{\begin{tabular}[c]{@{}c@{}}Eval. Scale\\ (\textbf{S}mall, \textbf{M}edium)\end{tabular}} \\ \midrule
\multirow{3}{*}{\textbf{Lighting}} & Gleam~\cite{Prakash2019-gb} & N          &  1     & Y                     & S   & L(2)         & Y(30)     & M(4)       \\
                                       & Xihe~\cite{Zhao2021-mg}       & Y(Replay) & 1   & Y & T & L(1) & N & S(2) \\
                                       & LitAR~\cite{Zhao2022-yx}      & Y(Game Engine)          & 1   & Y & S,T   & L(3) & N & S(1) \\
                                       \midrule
\multirow{2}{*}{\textbf{Depth}}                 & InDepth~\cite{Zhang2022-lq}   &  Y(Dataset)           &   1       & Y &  S        &   M(3/20)      &  Y(27) &  S(2)     \\
                                       & MobiDepth~\cite{Zhang2022-gg}  &  N          & 2       & Y*  &  S        &  -       &  N &  S(3)    \\
                                       \midrule
\multirow{3}{*}{\textbf{Tracking}}              & EdgeSLAM~\cite{Ben_Ali2022-jx}   &   N         &  1        &  Y*  &  S,T        &  M(4)       &  N  &  S(2)     \\
                                       & AdaptSLAM~\cite{Chen23AdaptSLAM}  &  Y          &  1         &  N(Dataset) &     S,T     &   M(6)      &  N  & S(1)       \\
                                       & FollowupAR~\cite{Xu2021-sn} &      N      &   1       & Y*  &    S      &  M(2/variations)       & N  &  S(1)     \\
                                       \midrule
\multirow{2}{*}{\begin{tabular}[c]{@{}l@{}}\textbf{Recognition/}\\ \textbf{ Detection}\end{tabular}} & CollabAR~\cite{Liu2020-fy}   &  Y(Dataset)          & 1 & Y*  &   S       &   M     & N  &  S(3)    \\
                                       & DeepMix~\cite{Guan2022-gv}    &   Y(Dataset)           &     1     &  Y*  &  S        &  L(-/3)      & Y(33)  &  S(1)     \\
                                       \midrule
\textbf{Scheduling}                             & Heimdall~\cite{Yi2020-na}  &      N      & K & Y  &   T   &  -        & N  &   S(2)    \\ \midrule
\textbf{Multi-User}                             & SEAR~\cite{Zhang2022-ml}  &    Y     & K & Y  &   T   &  -        & N  &   S(1)    \\ 
\bottomrule
\end{tabular}%
}
\vspace{-3mm}
\end{table*}

To understand the current practices and the limitations of AR evaluation, we surveyed 12 AR system papers focusing on their evaluation methodology. We categorize these papers based on AR tasks and characterize each paper's evaluation methodology along multiple important dimensions.
The dimensions are selected based on our prior experiences in evaluating AR systems. Table~\ref{tab:AR_sys_papers_eval} summarizes our findings where the \emph{Task(s)} column refers to the number of tasks each paper evaluated. 
This analysis presents an in-depth understanding of the AR researcher's evaluation workflow. We make several key observations.

First, most works require capturing visual data during experimentation; some even used specialized hardware to gather ground truth~\cite{Zhang2022-gg,Xu2021-sn,Guan2022-gv,Ben_Ali2022-jx,Liu2020-fy}. 
Specialized hardware can be costly, e.g., Microsoft HoloLens 2 used by DeepMix costs \$3.5K. Therefore, even if it is beneficial for evaluating AR systems on specialized hardware, not all papers can do so. 
However, if we can amortize the monetary cost by sharing the specialized devices among AR researchers, then it becomes more tractable to evaluate with specialized hardware. This suggests the need for an experimentation platform to provide the visual data-capturing feature and time-sharing of expensive specialized hardware. 

Second, many works evaluated and reported the performance with temporal and spatial variations. However, there is often a lack of \emph{explicit control} of the physical environment. For example, in MobiDepth~\cite{Zhang2022-gg}, the impact of spatial variance was measured by creating a dynamic scene that requires either moving the capturing device or the object of interest. However, the movement speed was only an approximated range, e.g., moving slowly vs. moving quickly. Although it was not explicitly mentioned, we suspect that the authors manually captured the required data. As such, it is hard to accurately quantify the impact of spatial locations on the AR systems (the proposed and the baselines). Temporal variations are slightly easier to control, e.g., in Xihe, we studied the impact of light intensity by fixing the rendering position and using a remotely controlled light source~\cite{Zhao2021-mg}. However, comparing how different systems work under the same temporal variations is non-trivial. We developed a session recorder to ensure consistent input to different systems~\cite{Zhao2021-mg}.

Third, we saw that almost all papers have low scene diversity and small-scale evaluation scenarios. For example, Xihe was only evaluated in one physical room to understand its real-world performance~\cite{Zhao2021-mg}. Even for works with higher scene diversity, they were only evaluated in up to six scenes~\cite{Chen23AdaptSLAM}. Additionally, most works evaluated use one to two mobile devices, with the upper end being four. Because of the heterogeneous mobile capabilities, it is hard to generalize and understand how well these works will perform in the wild. The requirements of having access to physical experiment spaces and high manual setup efforts seem to place a high toll on conducting diverse and large-scale evaluations.

Fourth, the nature of AR research calls for visual perception studies. All papers presented metric-based qualitative evaluations but only some conducted user studies to understand human perception performance.
User studies are often considered to have a higher barrier for entry, e.g., requiring researchers to recruit and manage participants and design scalable study protocols.

Last, almost all papers focus on single-task evaluation. 
However, delivering the AR experience from sensory data to rendered results to end users involve a complex ecosystem and many moving pieces (see Figure \ref{fig:ar_generic_pipeline}). For example, an AR shopping app often requires hand tracking and object detection models to allow users to interact with the mixed environment~\cite{Yi2020-na}. Furthermore, in the context of supporting multi-user AR experiences~\cite{Zhang2022-ml,Liu2020-fy}, it is unavoidable to consider inter-task dependencies. Evaluating one task in isolation is a good start, but we believe it would be better to evaluate the proposed solution in the context of the AR ecosystem. Such \emph{holistic evaluations} can provide valuable insights into how well the proposed solution will work in a real-world deployment.

\begin{figure*}[t]
    \centering
    \begin{subfigure}[b]{\columnwidth}
        \centering
        \includegraphics[width=\linewidth]{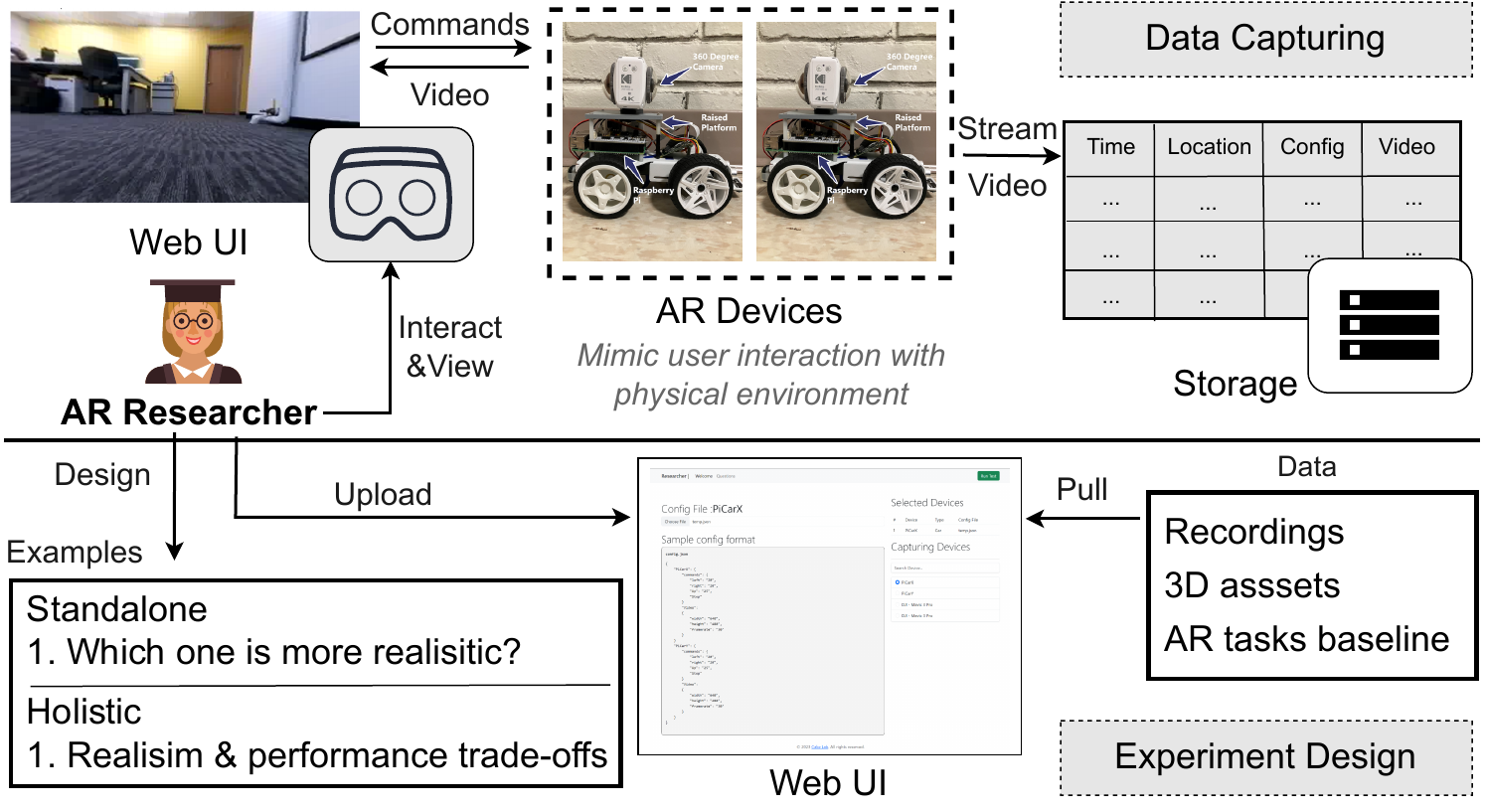}
        \caption{AR researcher}
        \label{fig:design_overview_researchers}
    \end{subfigure}%
    \quad
    \begin{subfigure}[b]{\columnwidth}
        \centering
        \includegraphics[width=\linewidth]{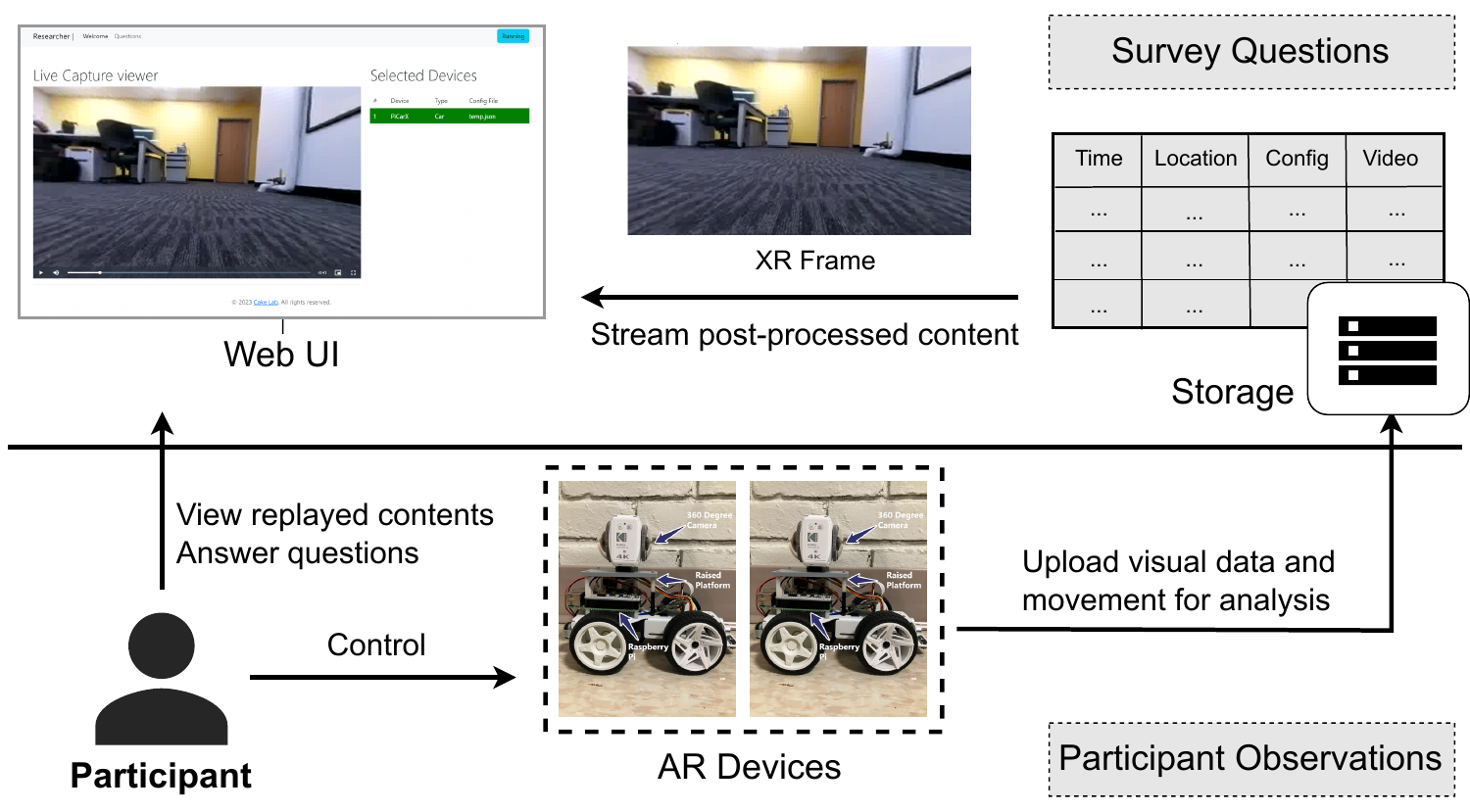}
        \caption{AR user study participant}
        \label{fig:design_overview_users}
    \end{subfigure}
    \vspace{-.7em}
    \caption{
       An overview of \sys key workflows. 
       \textnormal{AR researchers and user study participants can leverage \sys to perform key evaluation tasks, including data capturing, experiment design, participant observation, and online surveys.} 
    }
    \vspace{-1em}
    \label{fig:design_overview}
\end{figure*}

\vspace{-3mm}
\subsection{Reflections of Our Experimentation}

We re-examine what we did when evaluating an AR system  LitAR~\cite{Zhao2022-yx}. We used two groups of questions (Appendix~\ref{sec:reflection_qs}); the first group summarizes our evaluation process and rationales, while the second group asks for desirable features that can enable better evaluation experiences. 
We summarize the key reflection takeaways below.

\para{Controlled evaluation is time-consuming and difficult to set up.} Because AR evaluation often involves interaction with the physical environment, we need to control relevant physical factors during the experiments. For example, in LitAR,
we need to control the \emph{distance between observation and rendering positions} for each experiment run when evaluating its impact on lighting reconstruction quality. In the simplest case, this would involve manually setting up the capturing device and the props (i.e., a metal sphere ball) at specified locations and repeating the process for different distance variables. Each setup can take a few minutes, and therefore it can take many hours to complete a set of evaluations. However, this simple setup does not control other scene properties, e.g., lighting conditions and moving objects. Building a fully controlled physical scene is difficult (but not impossible). Instead, we resorted to a photorealistic indoor simulator that took about two months to set up.  

\para{Evaluation diversity and scale are limited by monetary cost and time.} When it comes to input data diversity, we are often limited by whatever off-the-shelf sensory devices are available and the budget to acquire them. For example, it would be interesting to see how different LiDAR sensors impact LitAR's performance, but one such device (e.g., iPad Pro) costs more than \$1K. Evaluating LitAR in more than three physical scene setups would be beneficial. Still, we were constrained by access to physical spaces and the ability to re-organize the scene (e.g., we don't have much control when using a public space). We also found ourselves developing \emph{one-off} tools and workflows when capturing various sensory data, streaming these data to the edge, and managing these data for further analysis. Though this is a similar challenge to edge offloading, the added burdens of dealing with hardware sensors and different AR tasks make it non-trivial to scale up the data capturing and, thus, evaluations. 

\para{Comparative studies are often guided by the easiness of reproducibility.} In LitAR, we only compared with two easy-to-setup baselines: a commercial solution ARKit and our prior work Xihe~\cite{Zhao2021-mg}. Would we benefit from comparing LitAR to other baselines? Probably. But such attempts were squashed by the hurdles to reproduce without source code and datasets or even the tremendous efforts required to set up the baselines. These hurdles also apply to user studies, which have other challenges, including participant recruitment~\cite{Zhang2022-lq} and multi-user coordination~\cite{Tung2021-uj}. Additionally, we only evaluated how LitAR performs for the lighting estimation task; we did not evaluate LitAR in an AR application to understand how it interacts with other AR tasks and its impact on the AR experiences. Even though this type of holistic evaluation is valuable for understanding in-the-wild performance, we see very few works that include holistic evaluations. We suspect that the lack of holistic evaluations is not caused by a lack of interest, but rather the lack of \emph{plug-and-play} evaluation support. It is already difficult to set up baselines for a single task; we can't imagine the obstacles one has to overcome to configure an entire application scenario that requires the coordination of many tasks.

Besides these three key observations, we suspect the fundamental problem that limits the AR evaluation methodology is \emph{treating evaluation as an afterthought}. In essence, we first design and build AR solutions and at a later stage, very reluctantly start the evaluation. The reluctance in part can be caused by the abovementioned challenges and obstacles in commencing any evaluations. \sys aims to simplify the evaluation process for AR solutions and promotes the \emph{key principle of iteration, prototyping, and testing}.

\vspace{-3mm}
\section{\sys Design}

This section describes our overall vision of \sys, a fully controllable and programmable AR platform, and its high-level design goals.
\sys is envisioned to operate as a standalone deployment (which we will describe an initial prototype in \S\ref{sec:impl}) or a federated platform that consists of network-connected deployments at different physical locations, similar to platforms such as PlanetLab and CloudLab~\cite{PlanetLab,cloudlab}.

\begin{figure}[t]
    \centering
    \includegraphics[width = 0.4 \textwidth]{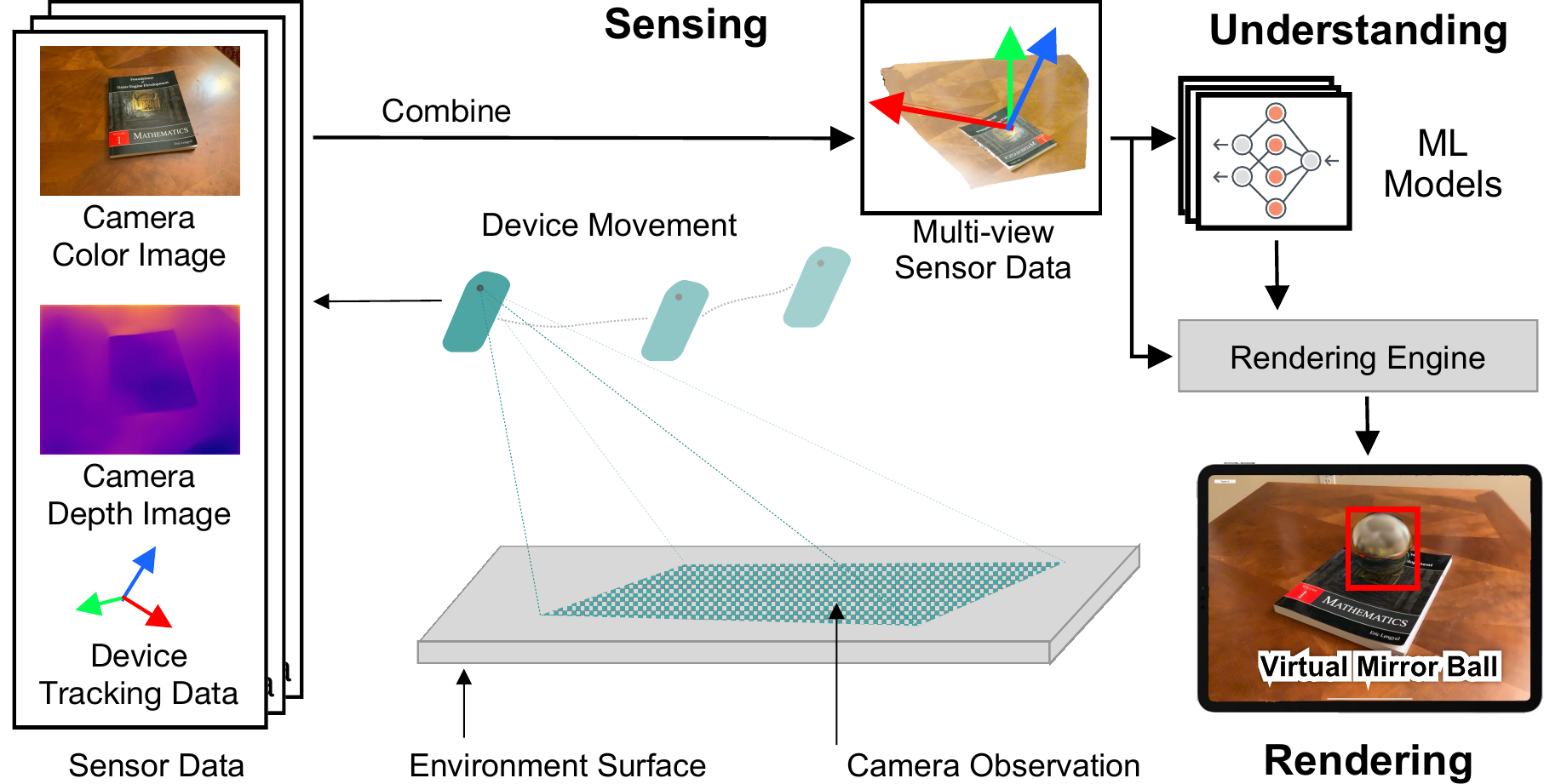}
    \vspace{-.5em}
    \caption{The generic sensing-understanding-rendering pipeline for AR. 
    \textnormal{
        Sensing leverages the increasingly rich sensors, understanding relies on compute-intensive DL models, and rendering overlays augmented information using modern rendering engines.}
    }
    \label{fig:ar_generic_pipeline}
    \vspace{-1em}
\end{figure}

\para{Key design insight:} \emph{let's generalize the AR pipeline!}
Based on our experiences in designing AR systems and surveying other works (see Table~\ref{tab:AR_sys_papers_eval}), we present a generic \emph{sensing-understanding-rendering} paradigm to which AR systems can generalize (Figure~\ref{fig:ar_generic_pipeline}).
Data capturing, i.e., \emph{sensing}, plays a critical role in using deep learning models to \emph{understand} the interaction between physical and virtual worlds. 
Those two components converge in the \emph{rendering} in which virtual objects are composited and shown to the end users~\cite{Du2020-ep,watson-2023-implicit-depth}.
By decomposing the AR pipeline into these three components, \sys can allow better sharing of each component among different active experiments and support holistic evaluations with minimal efforts from the AR researchers.

\para{Design goals.} We leverage our findings described in \S\ref{sec:motivation} and design \sys with the following three key goals. 
\1 \emph{Controllable evaluation environment}. \sys should provide the ability to control each physical scene programmatically. This includes but is not limited to controlling physical environment conditions such as lighting and object placements and data capturing devices. 
\2 \emph{Scalable data capturing and parallel evaluations.} \sys should allow both AR researchers and users to access a wide variety of hardware devices, e.g., to capture data in parallel from devices residing in different physical locations. Devices can be time-shared, and different user study participants can use different devices to enable truly large-scale evaluations.
\3  \emph{Reusable pipeline components and composable application scenarios.} \sys should provide built-in and default methods for different pipeline components. AR researchers can use existing components to define and configure their evaluation pipelines. To boost the component diversity, \sys will also allow community contribution, similar to existing AI/ML platforms like Hugging Face~\cite{huggingface}. Moreover, \sys will provide default application templates that AR researchers can drag and drop their tasks into, as well as the ability to customize the templates for easy-to-setup holistic evaluations.

\para{Overview.}
Figure~\ref{fig:design_overview} illustrates the key design components of a single deployment \sys and how two stakeholders, i.e., AR researchers and user study participants, interact with \sys. In a federated deployment, we will support a third stakeholder, the platform admin, who manages the operation of \sys. Additional example workflows are in Appendix~\ref{sec:example_workflows}.

\begin{figure}[t]
    \centering
    \vspace{-1em}
    \includegraphics[width = 0.47 \textwidth]{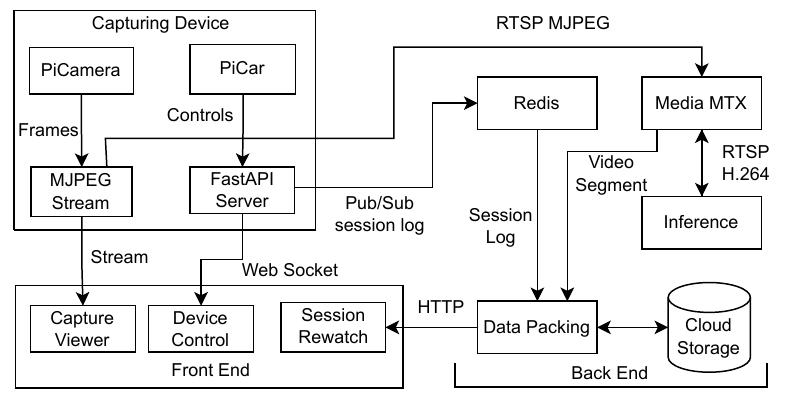}
    \vspace{-1em}
    \caption{
        A prototype implementation. 
        \textnormal{The capturing device is based on Raspberry Pi 3, and the backend runs in Google Cloud as microservices inside docker containers. 
        }
        }
    \vspace{-2em}
    \label{fig:implementation}
\end{figure}

The basic setup of \sys consists of \emph{data capturing devices}, a physical indoor scene where those capturing devices are initially parked and will explore, and a backend that persists data, including captured experimental data, static assets, and models/algorithms for different AR tasks. 
Many off-the-shelf hardware devices can act as the capturing devices, provided these devices are mobile, can be controlled remotely, and are equipped with necessary sensors, e.g., $360^{\circ}$ RGB cameras and depth sensors.
Example capturing devices include remote control cars and low-cost robots ~\cite{Liu2023ICRA,pyrobot2019,locobot}.

At a high level, AR researchers will programmatically control the capturing devices to traverse the physical scene and collect data helpful for understanding the environment, e.g., $360^{\circ}$ videos.
These data can be live viewed for monitoring and debugging purpose but will also be saved to facilitate future experiments. 
When an AR researcher needs to use \sys to capture the initial environment data, she can use any supported clients, e.g., VR headset or web UI, to select the desired physical scene. 
The scene configuration, which includes the number of capturing devices, capturing device capability, and the scene setup, like where the table is, is fixed for a given time. 
AR researchers select suitable scenes based on their experiment needs.
Note that AR researchers do not need to have physical access to each scene and do not need to own any of the physical resources. 
AR researchers are the users of \sys instead of the owners. 
This design shares the same spirit as existing experimental platforms such as PlanetLab, CloudLab, and AWS's device farm~\cite{aws_device_farm,PlanetLab,cloudlab}.

Once researchers finish setting up the experiments, e.g., capturing required data and configuring the AR pipeline with desired template and components, user study participants can leverage \sys to conduct large-scale online surveys or remotely use the AR pipeline for participant observation.

\section{Prototype Implementation}
\label{sec:impl}

The current prototype of \sys is implemented in Python and Javascript in a containerized microservice architecture.
Figure~\ref{fig:implementation} presents an overview of \sys's key components and their interactions. The prototype consists of three logical modules: the front end, the capturing device, and the back end. 
The \emph{front-end} module's primary objective is to provide \sys users, e.g., AR researchers and AR users, the ability to perform data capturing and monitor the capturing progress, as well as participate in the online survey via session rewatch.  
The \emph{capturing device} module, encapsulating the hardware, is responsible for data collection and dispatching the collected data to the \emph{back end}, which stores and processes the captured data. 
Currently, we implemented the front end in Vue.js, the data capturing based on a customized remote-controlled car PiCar-X~\cite{picarx} and a Raspberry Pi (RPi) 3B+, and the backend in containerized microservices running inside Google Cloud Platform (GCP)'s NVIDIA T4 GPU servers.
See Appendix~\ref{sec:module_level_impl_details} for per-component implementations.

\section{Preliminary Evaluation}

\begin{figure}[t]
    \centering
    \begin{subfigure}[b]{0.20\textwidth}
        \centering
        \includegraphics[width=\textwidth]{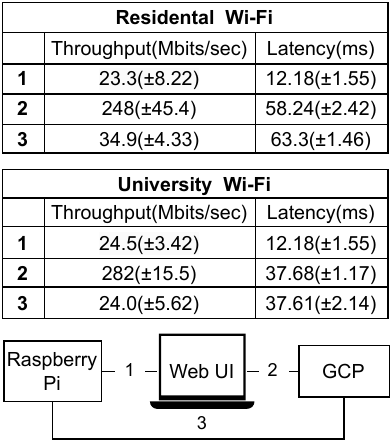}
        \caption{Setup}
        \label{fig:net-condition}
    \end{subfigure}
    \begin{subfigure}[b]{0.27\textwidth}
        \centering
        \vspace*{\fill}
        \includegraphics[width=\textwidth]{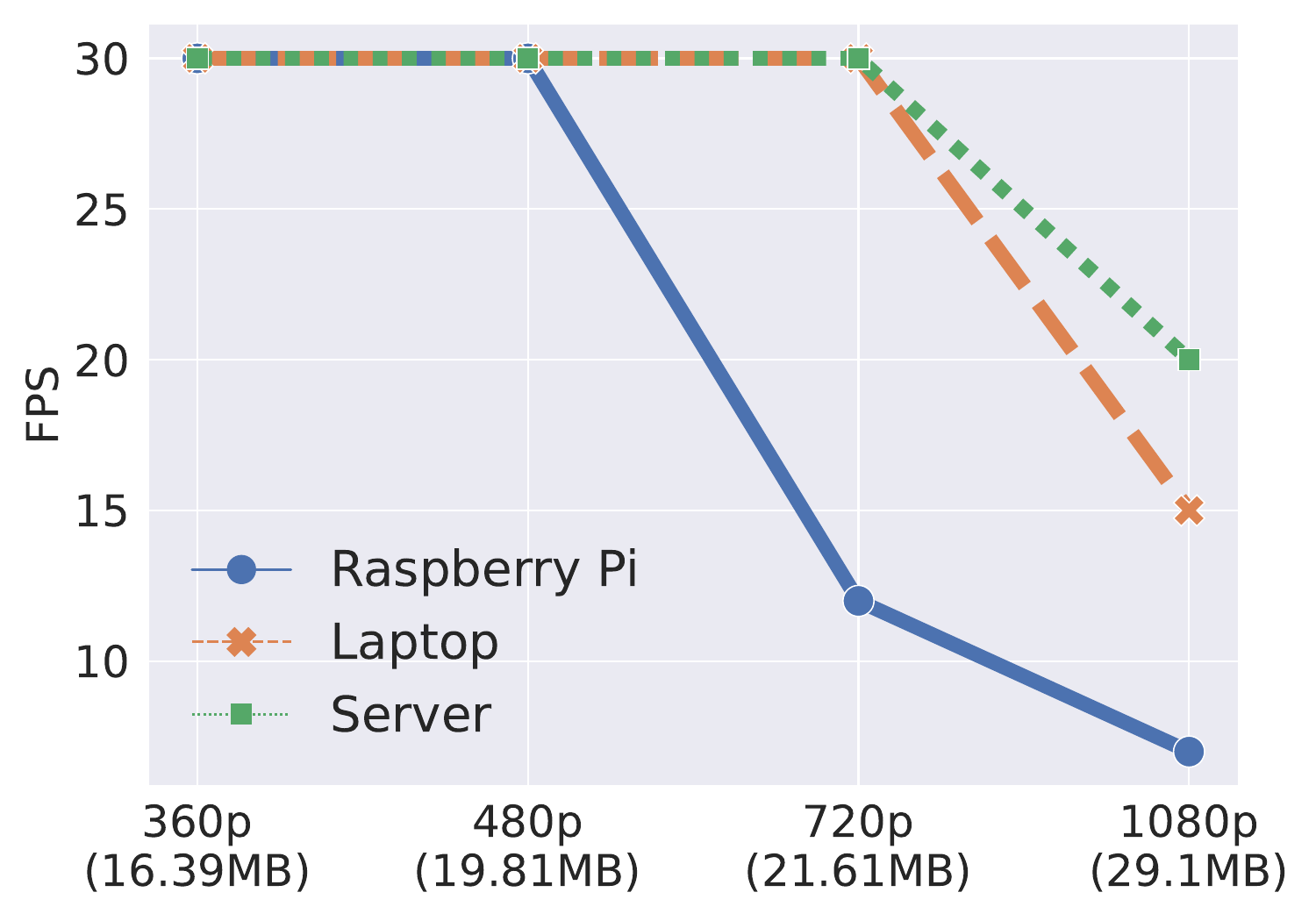}        
        \vspace*{\fill}
        \caption{Streaming frame rate}
        \label{fig:fps}
    \end{subfigure}
    \vspace{-.5em}
    \caption{
        Setup and streaming performance comparisons.
        \textnormal{Figure~\ref{fig:fps} shows \sys's FPS performance under the residential Wi-Fi\protect\footnotemark.}
    }
    \vspace{-3mm}
    \label{fig:network-fps}
\end{figure}

\footnotetext{We observe similar FPS trends in university Wi-Fi.}

We evaluate a prototype of \sys using a $360^{\circ}$ camera, three representative computation devices, and a back end deployed to the Google Cloud Platform (GCP), to understand the streaming performance.
The devices include:
\1 a lower-end Raspberry Pi 3, equipped with a 1.2GHz 64-bit quad-core ARM Cortex-A57 CPU with 1GB RAM;
\2 a mid-end laptop, powered by an Intel Core i7-11370H processor with 16GB RAM;
and \3 a high-end server with a 4th Generation Intel Xeon processor and comes with 64GB RAM. 
Figure~\ref{fig:net-condition} depicts the setup and the pairwise network performance measured using \texttt{iPerf} under two Wi-Fi networks. 

Figure~\ref{fig:fps} compares the frame per second (FPS) achieved under different streaming conditions. 
For each condition, we vary the frame resolution and the streaming device pairs.
A streaming pair consists of one of the three devices (Raspberry Pi, laptop, and server) and an endpoint GCP server. 
Each device will stream the same video, recorded with the $360^{\circ}$ camera, from a file. 
We make two key observations.
First, as the resolution increases from 360p to 1080p, the FPS decreases for all devices.
This suggests that the network starts to become the bottleneck (by comparing the network bandwidth to the streaming data size) and the device can't keep up with streaming larger $360^{\circ}$ frames.
Second, even the high-end server cannot achieve the desired 30 FPS when streaming at 1080p. We suspect that our CPU-based implementation is the culprit to such performance. 
We monitor the resource utilization and find that CPU utilizations increase drastically for larger resolutions (details in  Appendix~\ref{subsec:additional_results:resourceutil}).
In short, our preliminary evaluations suggest the need to further investigate network optimizations and GPU-based implementations for supporting high-quality experimentation data streaming.

\begin{table}[t]
\centering
\caption{
Task latency comparison over ten runs.
}
\vspace{-.5em}
\label{tab:metrics}
\resizebox{0.9\columnwidth}{!}{%
\begin{tabular}{@{}l|rr@{}}
\toprule
\textbf{Task} & \textbf{Residential Wi-Fi} & \textbf{University Wi-Fi} \\ \midrule
Loading the system & 222.49 ($\pm$ 8.55 ms) & 134.33 ($\pm$ 3.93ms) \\ 
Setting up the device & 68.40 ($\pm$ 8.42 ms) & 41.76 ($\pm$ 2.53ms)\\ \midrule
Client-server latency & 63.30 ($\pm$1.46 ms) & 38.06($\pm$ 2.24ms) \\ 
Executing control command & 6.11 ($\pm$ 1.81 ms) & 1.81 ($\pm$ 0.77ms)\\ 
\bottomrule
\end{tabular}
}
\vspace{-5mm}
\end{table}

To better understand how \sys operates, we also evaluate the per-task latency under two Wi-Fi conditions.
Table~\ref{tab:metrics} summarizes the per-task latency:
\1 \emph{loading the system:} is the amount of time takes to get a response from our GCP Redis container.
\2 \emph{setting up the device:} is the amount of time takes to write the device ID to the Redis database,
\3 \emph{client-server latency:} describes the time to confirm that the server is active by pinging the server,
\4 and \emph{executing commands:} is the amount of time takes for the PiCar-X to execute the user's commands.
We find that it takes 176.09 ms and 290.89 ms to start up \sys using university and residential Wi-Fi respectively. We believe this overhead is reasonable because it is a one-time setup. 
The latency between client-server can be as high as 63.3 ms and can cause performance issues under certain interactive evaluation tasks. This suggests the need for a geo-distributed \sys deployment and considering the server locations based on \sys users' locations. 
University Wi-Fi in general leads to lower task latency and suggests the importance of properly configuring the network when using \sys for interactive experimentation.

\section{Conclusion}

This paper describes the design of \sys, an AR experimentation platform that allows setting up controllable evaluation environments easily, capturing data scalably, conducting evaluations in parallel, and reusing evaluation components. 
Our design is based on an in-depth analysis of the evaluation methodology from 12 recent AR system papers and our prior experiences and centers around the key insight of generalizable AR pipelines. 
\sys can allow AR researchers to share the physical devices and physical spaces, increasing the evaluation scale and diversity currently lacking.
A prototype implementation and preliminary evaluation \sys were also presented, revealing interesting future directions such as improving the capturing device's onboard processing power and optimizing the network performance between the front end and the capturing device. 
We will iterate the design and implementation of a local deployment while working on our AR projects. 
We hope to provide \sys as a service to the research community.

\begin{acks}
\end{acks}
This work was supported in part by NSF Grants \#2105564 and \#2236987, a VMware grant, the Worcester Polytechnic Institute’s Computer Science Department, and Google Cloud Services. We thank the MQP team, consisting of Greg V. Klimov, Samuel L. Kwok, Cole R. Manning, Mason V. Powell, Sam P. Rowe, and Adam Yang, for their efforts in assembling and testing the RC-based capturing device, and Maya Angeles' help with hardware design.

\clearpage
\balance
\bibliographystyle{ACM-Reference-Format}
\bibliography{ImmerCom}


\begin{thebibliography}{32}


\ifx \showCODEN    \undefined \def \showCODEN     #1{\unskip}     \fi
\ifx \showDOI      \undefined \def \showDOI       #1{#1}\fi
\ifx \showISBNx    \undefined \def \showISBNx     #1{\unskip}     \fi
\ifx \showISBNxiii \undefined \def \showISBNxiii  #1{\unskip}     \fi
\ifx \showISSN     \undefined \def \showISSN      #1{\unskip}     \fi
\ifx \showLCCN     \undefined \def \showLCCN      #1{\unskip}     \fi
\ifx \shownote     \undefined \def \shownote      #1{#1}          \fi
\ifx \showarticletitle \undefined \def \showarticletitle #1{#1}   \fi
\ifx \showURL      \undefined \def \showURL       {\relax}        \fi
\providecommand\bibfield[2]{#2}
\providecommand\bibinfo[2]{#2}
\providecommand\natexlab[1]{#1}
\providecommand\showeprint[2][]{arXiv:#2}

\bibitem[Apicharttrisorn et~al\mbox{.}(2023)]%
        {Apicharttrisorn2022Sensys}
\bibfield{author}{\bibinfo{person}{Kittipat Apicharttrisorn},
  \bibinfo{person}{Jiasi Chen}, \bibinfo{person}{Vyas Sekar},
  \bibinfo{person}{Anthony Rowe}, {and} \bibinfo{person}{Srikanth~V.
  Krishnamurthy}.} \bibinfo{year}{2023}\natexlab{}.
\newblock \showarticletitle{Breaking Edge Shackles: Infrastructure-Free
  Collaborative Mobile Augmented Reality} \emph{(\bibinfo{series}{SenSys
  '22})}. \bibinfo{publisher}{Association for Computing Machinery},
  \bibinfo{address}{New York, NY, USA}, \bibinfo{pages}{1–15}.
\newblock
\showISBNx{9781450398862}
\urldef\tempurl%
\url{https://doi.org/10.1145/3560905.3568546}
\showDOI{\tempurl}


\bibitem[Ben~Ali et~al\mbox{.}(2022)]%
        {Ben_Ali2022-jx}
\bibfield{author}{\bibinfo{person}{Ali~J Ben~Ali}, \bibinfo{person}{Marziye
  Kouroshli}, \bibinfo{person}{Sofiya Semenova}, \bibinfo{person}{Zakieh~Sadat
  Hashemifar}, \bibinfo{person}{Steven~Y Ko}, {and} \bibinfo{person}{Karthik
  Dantu}.} \bibinfo{year}{2022}\natexlab{}.
\newblock \showarticletitle{{Edge-SLAM: Edge-Assisted Visual Simultaneous
  Localization and Mapping}}.
\newblock \bibinfo{journal}{\emph{ACM Trans. Embed. Comput. Syst.}}
  \bibinfo{volume}{22}, \bibinfo{number}{1} (\bibinfo{date}{Oct.}
  \bibinfo{year}{2022}), \bibinfo{pages}{1--31}.
\newblock


\bibitem[Chen et~al\mbox{.}(2023)]%
        {Chen23AdaptSLAM}
\bibfield{author}{\bibinfo{person}{Ying Chen}, \bibinfo{person}{Hazer
  Inaltekin}, {and} \bibinfo{person}{Maria Gorlatova}.}
  \bibinfo{year}{2023}\natexlab{}.
\newblock \showarticletitle{{AdaptSLAM}: Edge-Assisted Adaptive SLAM with
  Resource Constraints via Uncertainty Minimization}. In
  \bibinfo{booktitle}{\emph{Proc. IEEE INFOCOM}}.
\newblock


\bibitem[Du et~al\mbox{.}(2020)]%
        {Du2020-ep}
\bibfield{author}{\bibinfo{person}{Ruofei Du}, \bibinfo{person}{Eric Turner},
  \bibinfo{person}{Maksym Dzitsiuk}, \bibinfo{person}{Luca Prasso},
  \bibinfo{person}{Ivo Duarte}, \bibinfo{person}{Jason Dourgarian},
  \bibinfo{person}{Joao Afonso}, \bibinfo{person}{Jose Pascoal},
  \bibinfo{person}{Josh Gladstone}, \bibinfo{person}{Nuno Cruces},
  \bibinfo{person}{Shahram Izadi}, \bibinfo{person}{Adarsh Kowdle},
  \bibinfo{person}{Konstantine Tsotsos}, {and} \bibinfo{person}{David Kim}.}
  \bibinfo{year}{2020}\natexlab{}.
\newblock \showarticletitle{{DepthLab: Real-time 3D Interaction with Depth Maps
  for Mobile Augmented Reality}}. In \bibinfo{booktitle}{\emph{{Proceedings of
  the 33rd Annual ACM Symposium on User Interface Software and Technology}}}
  \emph{(\bibinfo{series}{UIST '20})}.
\newblock


\bibitem[Duplyakin et~al\mbox{.}(2019)]%
        {cloudlab}
\bibfield{author}{\bibinfo{person}{Dmitry Duplyakin}, \bibinfo{person}{Robert
  Ricci}, \bibinfo{person}{Aleksander Maricq}, \bibinfo{person}{Gary Wong},
  \bibinfo{person}{Jonathon Duerig}, \bibinfo{person}{Eric Eide},
  \bibinfo{person}{Leigh Stoller}, \bibinfo{person}{Mike Hibler},
  \bibinfo{person}{David Johnson}, \bibinfo{person}{Kirk Webb},
  \bibinfo{person}{Aditya Akella}, \bibinfo{person}{Kuangching Wang},
  \bibinfo{person}{Glenn Ricart}, \bibinfo{person}{Larry Landweber},
  \bibinfo{person}{Chip Elliott}, \bibinfo{person}{Michael Zink},
  \bibinfo{person}{Emmanuel Cecchet}, \bibinfo{person}{Snigdhaswin Kar}, {and}
  \bibinfo{person}{Prabodh Mishra}.} \bibinfo{year}{2019}\natexlab{}.
\newblock \showarticletitle{The Design and Operation of {CloudLab}}. In
  \bibinfo{booktitle}{\emph{Proceedings of the {USENIX} Annual Technical
  Conference (ATC)}}. \bibinfo{pages}{1--14}.
\newblock


\bibitem[Experience(2022)]%
        {green_planet_ar}
\bibfield{author}{\bibinfo{person}{The Green Planet~AR Experience}.}
  \bibinfo{year}{2022}\natexlab{}.
\newblock
  \bibinfo{howpublished}{\url{https://www.factory42.uk/greenplanetexperience}}.
\newblock


\bibitem[Face(2023)]%
        {huggingface}
\bibfield{author}{\bibinfo{person}{Hugging Face}.}
  \bibinfo{year}{2023}\natexlab{}.
\newblock \bibinfo{howpublished}{\url{https://huggingface.co/}}.
\newblock


\bibitem[Farm(2023)]%
        {aws_device_farm}
\bibfield{author}{\bibinfo{person}{AWS~Device Farm}.}
  \bibinfo{year}{2023}\natexlab{}.
\newblock \bibinfo{howpublished}{\url{https://aws.amazon.com/device-farm/}}.
\newblock


\bibitem[for XR~(Preview)(2023)]%
        {googlexr}
\bibfield{author}{\bibinfo{person}{Immersive~Stream for XR~(Preview)}.}
  \bibinfo{year}{2023}\natexlab{}.
\newblock \bibinfo{howpublished}{\url{https://xr.withgoogle.com/}}.
\newblock


\bibitem[Guan et~al\mbox{.}(2022)]%
        {Guan2022-gv}
\bibfield{author}{\bibinfo{person}{Yongjie Guan}, \bibinfo{person}{Xueyu Hou},
  \bibinfo{person}{Nan Wu}, \bibinfo{person}{Bo Han}, {and}
  \bibinfo{person}{Tao Han}.} \bibinfo{year}{2022}\natexlab{}.
\newblock \showarticletitle{{DeepMix: mobility-aware, lightweight, and hybrid
  3D object detection for headsets}}. In \bibinfo{booktitle}{\emph{{Proceedings
  of the 20th Annual International Conference on Mobile Systems, Applications
  and Services}}} \emph{(\bibinfo{series}{MobiSys '22})}.
  \bibinfo{pages}{28--41}.
\newblock


\bibitem[Han et~al\mbox{.}(2022)]%
        {Han2022-bh}
\bibfield{author}{\bibinfo{person}{Bo Han}, \bibinfo{person}{Parth Pathak},
  \bibinfo{person}{Songqing Chen}, {and} \bibinfo{person}{Lap-Fai~Craig Yu}.}
  \bibinfo{year}{2022}\natexlab{}.
\newblock \showarticletitle{{CoMIC: A Collaborative Mobile Immersive Computing
  Infrastructure for Conducting Multi-user XR Research}}.
\newblock \bibinfo{journal}{\emph{IEEE Netw.}} (\bibinfo{year}{2022}),
  \bibinfo{pages}{1--9}.
\newblock


\bibitem[Huzaifa et~al\mbox{.}(2021)]%
        {Huzaifa2021-jj}
\bibfield{author}{\bibinfo{person}{Muhammad Huzaifa}, \bibinfo{person}{Rishi
  Desai}, \bibinfo{person}{Samuel Grayson}, \bibinfo{person}{Xutao Jiang},
  \bibinfo{person}{Ying Jing}, \bibinfo{person}{Jae Lee}, \bibinfo{person}{Fang
  Lu}, \bibinfo{person}{Yihan Pang}, \bibinfo{person}{Joseph Ravichandran},
  \bibinfo{person}{Finn Sinclair}, \bibinfo{person}{Boyuan Tian},
  \bibinfo{person}{Hengzhi Yuan}, \bibinfo{person}{Jeffrey Zhang}, {and}
  \bibinfo{person}{Sarita~V Adve}.} \bibinfo{year}{2021}\natexlab{}.
\newblock \showarticletitle{{ILLIXR: Enabling End-to-End Extended Reality
  Research}}. In \bibinfo{booktitle}{\emph{{2021 IEEE International Symposium
  on Workload Characterization (IISWC)}}}. \bibinfo{pages}{24--38}.
\newblock


\bibitem[Kwon et~al\mbox{.}(2023)]%
        {Kwon2022-vm-mlsys}
\bibfield{author}{\bibinfo{person}{Hyoukjun Kwon},
  \bibinfo{person}{Krishnakumar Nair}, \bibinfo{person}{Jamin Seo},
  \bibinfo{person}{Jason Yik}, \bibinfo{person}{Debabrata Mohapatra},
  \bibinfo{person}{Dongyuan Zhan}, \bibinfo{person}{Jinook Song},
  \bibinfo{person}{Peter Capak}, \bibinfo{person}{Peizhao Zhang},
  \bibinfo{person}{Peter Vajda}, \bibinfo{person}{Colby Banbury},
  \bibinfo{person}{Mark Mazumder}, \bibinfo{person}{Liangzhen Lai},
  \bibinfo{person}{Ashish Sirasao}, \bibinfo{person}{Tushar Krishna},
  \bibinfo{person}{Harshit Khaitan}, \bibinfo{person}{Vikas Chandra}, {and}
  \bibinfo{person}{Vijay~Janapa Reddi}.} \bibinfo{year}{2023}\natexlab{}.
\newblock \showarticletitle{XRBench: An extended reality (XR) machine learning
  benchmark suite for the metaverse}. In \bibinfo{booktitle}{\emph{Proceedings
  of Machine Learning and Systems}}, Vol.~\bibinfo{volume}{5}.
\newblock


\bibitem[{Larry L. Peterson and Prof. David Culler}(2002)]%
        {PlanetLab}
\bibfield{author}{\bibinfo{person}{{Larry L. Peterson and Prof. David
  Culler}}.} \bibinfo{year}{2002}\natexlab{}.
\newblock \bibinfo{title}{{PlanetLab}}.
\newblock \bibinfo{howpublished}{\url{https://planetlab.cs.princeton.edu/}}.
\newblock


\bibitem[Liu et~al\mbox{.}(2023)]%
        {Liu2023ICRA}
\bibfield{author}{\bibinfo{person}{Liangkai Liu}, \bibinfo{person}{Ren Zhong},
  \bibinfo{person}{Aaron Willcock}, \bibinfo{person}{Nathan Fisher}, {and}
  \bibinfo{person}{Weisong Shi}.} \bibinfo{year}{2023}\natexlab{}.
\newblock \showarticletitle{An Open Approach to Energy-Efficient Autonomous
  Mobile Robots citation}. In \bibinfo{booktitle}{\emph{2023 International
  Conference on Robotics and Automation (ICRA)}}. \bibinfo{publisher}{IEEE
  Press}, \bibinfo{numpages}{7}~pages.
\newblock


\bibitem[Liu et~al\mbox{.}(2020)]%
        {Liu2020-fy}
\bibfield{author}{\bibinfo{person}{Z Liu}, \bibinfo{person}{G Lan},
  \bibinfo{person}{J Stojkovic}, \bibinfo{person}{Y Zhang}, \bibinfo{person}{C
  Joe-Wong}, {and} \bibinfo{person}{M Gorlatova}.}
  \bibinfo{year}{2020}\natexlab{}.
\newblock \showarticletitle{{CollabAR: Edge-assisted Collaborative Image
  Recognition for Mobile Augmented Reality}}. In
  \bibinfo{booktitle}{\emph{{2020 19th ACM/IEEE International Conference on
  Information Processing in Sensor Networks (IPSN)}}}.
  \bibinfo{pages}{301--312}.
\newblock


\bibitem[Mandlekar et~al\mbox{.}(2018)]%
        {roboturk}
\bibfield{author}{\bibinfo{person}{Ajay Mandlekar}, \bibinfo{person}{Yuke Zhu},
  \bibinfo{person}{Animesh Garg}, \bibinfo{person}{Jonathan Booher},
  \bibinfo{person}{Max Spero}, \bibinfo{person}{Albert Tung},
  \bibinfo{person}{Julian Gao}, \bibinfo{person}{John Emmons},
  \bibinfo{person}{Anchit Gupta}, \bibinfo{person}{Emre Orbay},
  \bibinfo{person}{Silvio Savarese}, {and} \bibinfo{person}{Li Fei-Fei}.}
  \bibinfo{year}{2018}\natexlab{}.
\newblock \showarticletitle{ROBOTURK: A Crowdsourcing Platform for Robotic
  Skill Learning through Imitation}. In \bibinfo{booktitle}{\emph{Conference on
  Robot Learning}}.
\newblock


\bibitem[Murali et~al\mbox{.}(2019)]%
        {pyrobot2019}
\bibfield{author}{\bibinfo{person}{Adithyavairavan Murali},
  \bibinfo{person}{Tao Chen}, \bibinfo{person}{Kalyan~Vasudev Alwala},
  \bibinfo{person}{Dhiraj Gandhi}, \bibinfo{person}{Lerrel Pinto},
  \bibinfo{person}{Saurabh Gupta}, {and} \bibinfo{person}{Abhinav Gupta}.}
  \bibinfo{year}{2019}\natexlab{}.
\newblock \showarticletitle{PyRobot: An Open-source Robotics Framework for
  Research and Benchmarking}.
\newblock \bibinfo{journal}{\emph{arXiv preprint arXiv:1906.08236}}
  (\bibinfo{year}{2019}).
\newblock


\bibitem[{Oliver Kroemer}(2023)]%
        {locobot}
\bibfield{author}{\bibinfo{person}{{Oliver Kroemer}}.}
  \bibinfo{year}{2023}\natexlab{}.
\newblock \bibinfo{title}{{LoCoBot: An Open Source Low Cost Robot}}.
\newblock \bibinfo{howpublished}{\url{http://www.locobot.org/}}.
\newblock


\bibitem[Prakash et~al\mbox{.}(2019)]%
        {Prakash2019-gb}
\bibfield{author}{\bibinfo{person}{Siddhant Prakash}, \bibinfo{person}{Alireza
  Bahremand}, \bibinfo{person}{Linda~D Nguyen}, {and} \bibinfo{person}{Robert
  LiKamWa}.} \bibinfo{year}{2019}\natexlab{}.
\newblock \showarticletitle{{GLEAM: An Illumination Estimation Framework for
  Real-time Photorealistic Augmented Reality on Mobile Devices}}. In
  \bibinfo{booktitle}{\emph{{Proceedings of the 17th Annual International
  Conference on Mobile Systems, Applications, and Services}}} (Seoul, Republic
  of Korea) \emph{(\bibinfo{series}{MobiSys '19})}.
  \bibinfo{publisher}{Association for Computing Machinery},
  \bibinfo{address}{New York, NY, USA}.
\newblock


\bibitem[SunFounder(2023)]%
        {picarx}
\bibfield{author}{\bibinfo{person}{SunFounder}.}
  \bibinfo{year}{2023}\natexlab{}.
\newblock \bibinfo{title}{{Raspberry Pi Ai Car Kit - PiCar-X}}.
\newblock
  \bibinfo{howpublished}{\url{https://www.sunfounder.com/products/picar-x}}.
\newblock
\newblock
\shownote{Accessed: 2023-6-15}.


\bibitem[Tung et~al\mbox{.}(2021)]%
        {Tung2021-uj}
\bibfield{author}{\bibinfo{person}{Albert Tung}, \bibinfo{person}{Josiah Wong},
  \bibinfo{person}{Ajay Mandlekar}, \bibinfo{person}{Roberto
  Mart{\'\i}n-Mart{\'\i}n}, \bibinfo{person}{Yuke Zhu}, \bibinfo{person}{Li
  Fei-Fei}, {and} \bibinfo{person}{Silvio Savarese}.}
  \bibinfo{year}{2021}\natexlab{}.
\newblock \showarticletitle{{Learning Multi-Arm Manipulation Through
  Collaborative Teleoperation}}. In \bibinfo{booktitle}{\emph{{2021 IEEE
  International Conference on Robotics and Automation (ICRA)}}}.
  \bibinfo{pages}{9212--9219}.
\newblock


\bibitem[Watson et~al\mbox{.}(2023)]%
        {watson-2023-implicit-depth}
\bibfield{author}{\bibinfo{person}{Jamie Watson}, \bibinfo{person}{Mohamed
  Sayed}, \bibinfo{person}{Zawar Qureshi}, \bibinfo{person}{Gabriel~J.
  Brostow}, \bibinfo{person}{Sara Vicente}, \bibinfo{person}{Oisin~Mac Aodha},
  {and} \bibinfo{person}{Michael Firman}.} \bibinfo{year}{2023}\natexlab{}.
\newblock \showarticletitle{Virtual Occlusions Through Implicit Depth}. In
  \bibinfo{booktitle}{\emph{CVPR}}.
\newblock


\bibitem[Xu et~al\mbox{.}(2021)]%
        {Xu2021-sn}
\bibfield{author}{\bibinfo{person}{Jingao Xu}, \bibinfo{person}{Guoxuan Chi},
  \bibinfo{person}{Zheng Yang}, \bibinfo{person}{Danyang Li},
  \bibinfo{person}{Qian Zhang}, \bibinfo{person}{Qiang Ma}, {and}
  \bibinfo{person}{Xin Miao}.} \bibinfo{year}{2021}\natexlab{}.
\newblock \showarticletitle{{FollowUpAR: enabling follow-up effects in mobile
  AR applications}}. In \bibinfo{booktitle}{\emph{{Proceedings of the 19th
  Annual International Conference on Mobile Systems, Applications, and
  Services}}} \emph{(\bibinfo{series}{MobiSys '21})}.
  \bibinfo{publisher}{Association for Computing Machinery},
  \bibinfo{pages}{1--13}.
\newblock


\bibitem[Yi and Lee(2020)]%
        {Yi2020-na}
\bibfield{author}{\bibinfo{person}{Juheon Yi} {and} \bibinfo{person}{Youngki
  Lee}.} \bibinfo{year}{2020}\natexlab{}.
\newblock \showarticletitle{{Heimdall: mobile GPU coordination platform for
  augmented reality applications}}. In \bibinfo{booktitle}{\emph{{Proceedings
  of the 26th Annual International Conference on Mobile Computing and
  Networking}}} (London, United Kingdom) \emph{(\bibinfo{series}{MobiCom '20},
  \bibinfo{number}{Article 35})}. \bibinfo{publisher}{Association for Computing
  Machinery}, \bibinfo{address}{New York, NY, USA}, \bibinfo{pages}{1--14}.
\newblock


\bibitem[Zhang et~al\mbox{.}(2022c)]%
        {Zhang2022-gg}
\bibfield{author}{\bibinfo{person}{Jinrui Zhang}, \bibinfo{person}{Huan Yang},
  \bibinfo{person}{Ju Ren}, \bibinfo{person}{Deyu Zhang},
  \bibinfo{person}{Bangwen He}, \bibinfo{person}{Ting Cao},
  \bibinfo{person}{Yuanchun Li}, \bibinfo{person}{Yaoxue Zhang}, {and}
  \bibinfo{person}{Yunxin Liu}.} \bibinfo{year}{2022}\natexlab{c}.
\newblock \showarticletitle{{MobiDepth: real-time depth estimation using
  on-device dual cameras}}. In \bibinfo{booktitle}{\emph{{Proceedings of the
  28th Annual International Conference on Mobile Computing And Networking}}}
  \emph{(\bibinfo{series}{MobiCom '22})}.
\newblock


\bibitem[Zhang et~al\mbox{.}(2022a)]%
        {Zhang2022-ml}
\bibfield{author}{\bibinfo{person}{Wenxiao Zhang}, \bibinfo{person}{Bo Han},
  {and} \bibinfo{person}{Pan Hui}.} \bibinfo{year}{2022}\natexlab{a}.
\newblock \showarticletitle{{SEAR: Scaling Experiences in Multi-user Augmented
  Reality}}.
\newblock \bibinfo{journal}{\emph{IEEE Trans. Vis. Comput. Graph.}}
  \bibinfo{volume}{28}, \bibinfo{number}{5} (\bibinfo{date}{May}
  \bibinfo{year}{2022}), \bibinfo{pages}{1982--1992}.
\newblock


\bibitem[Zhang et~al\mbox{.}(2022b)]%
        {Zhang2022-lq}
\bibfield{author}{\bibinfo{person}{Yunfan Zhang}, \bibinfo{person}{Tim
  Scargill}, \bibinfo{person}{Ashutosh Vaishnav}, \bibinfo{person}{Gopika
  Premsankar}, \bibinfo{person}{Mario Di~Francesco}, {and}
  \bibinfo{person}{Maria Gorlatova}.} \bibinfo{year}{2022}\natexlab{b}.
\newblock \showarticletitle{{InDepth: Real-time Depth Inpainting for Mobile
  Augmented Reality}}.
\newblock \bibinfo{journal}{\emph{Proc. ACM Interact. Mob. Wearable Ubiquitous
  Technol.}} \bibinfo{volume}{6}, \bibinfo{number}{1} (\bibinfo{date}{March}
  \bibinfo{year}{2022}), \bibinfo{pages}{1--25}.
\newblock


\bibitem[Zhao et~al\mbox{.}(2023)]%
        {zhao2023hotmobile}
\bibfield{author}{\bibinfo{person}{Yiqin Zhao}, \bibinfo{person}{Sean Fanello},
  {and} \bibinfo{person}{Tian Guo}.} \bibinfo{year}{2023}\natexlab{}.
\newblock \showarticletitle{Multi-Camera Lighting Estimation for Photorealistic
  Front-Facing Mobile Augmented Reality}. In
  \bibinfo{booktitle}{\emph{Proceedings of the 24th International Workshop on
  Mobile Computing Systems and Applications}} (Newport Beach, California)
  \emph{(\bibinfo{series}{HotMobile '23})}. \bibinfo{publisher}{Association for
  Computing Machinery}, \bibinfo{address}{New York, NY, USA},
  \bibinfo{pages}{68–73}.
\newblock
\showISBNx{9798400700170}
\urldef\tempurl%
\url{https://doi.org/10.1145/3572864.3580337}
\showDOI{\tempurl}


\bibitem[Zhao and Guo(2020)]%
        {Zhao2020-gr-eccv}
\bibfield{author}{\bibinfo{person}{Yiqin Zhao} {and} \bibinfo{person}{Tian
  Guo}.} \bibinfo{year}{2020}\natexlab{}.
\newblock \showarticletitle{{PointAR: Efficient Lighting Estimation for Mobile
  Augmented Reality}}. In \bibinfo{booktitle}{\emph{The European Conference on
  Computer Vision}} \emph{(\bibinfo{series}{ECCV '20})}.
  \bibinfo{pages}{678--693}.
\newblock


\bibitem[Zhao and Guo(2021)]%
        {Zhao2021-mg}
\bibfield{author}{\bibinfo{person}{Yiqin Zhao} {and} \bibinfo{person}{Tian
  Guo}.} \bibinfo{year}{2021}\natexlab{}.
\newblock \showarticletitle{{Xihe: a 3D vision-based lighting estimation
  framework for mobile augmented reality}}. In
  \bibinfo{booktitle}{\emph{{Proceedings of the 19th Annual International
  Conference on Mobile Systems, Applications, and Services}}}
  \emph{(\bibinfo{series}{MobiSys '21})}. \bibinfo{pages}{28--40}.
\newblock


\bibitem[Zhao et~al\mbox{.}(2022)]%
        {Zhao2022-yx}
\bibfield{author}{\bibinfo{person}{Yiqin Zhao}, \bibinfo{person}{Chongyang Ma},
  \bibinfo{person}{Haibin Huang}, {and} \bibinfo{person}{Tian Guo}.}
  \bibinfo{year}{2022}\natexlab{}.
\newblock \showarticletitle{{LITAR: Visually Coherent Lighting for Mobile
  Augmented Reality}}.
\newblock \bibinfo{journal}{\emph{Proc. ACM Interact. Mob. Wearable Ubiquitous
  Technol.}} \bibinfo{volume}{6}, \bibinfo{number}{3} (\bibinfo{date}{Sept.}
  \bibinfo{year}{2022}), \bibinfo{pages}{1--29}.
\newblock


\end{thebibliography}

\clearpage
\appendix
\section{\sys Example Workflows}
\label{sec:example_workflows}

\subsection{Design for AR Researchers}
\label{subsec:design_for_researchers}

We envision that \sys can aid AR developers and researchers in evaluating AR systems better by providing features including \emph{data capturing}, \emph{sensing visualization}, \emph{experimentation design}, and \emph{reproducible pipeline evaluation}. 
This section describes the design of \sys by explaining how an AR researcher will use \sys to accomplish the key evaluation tasks of \emph{data capturing} and \emph{experimentation design}.
Figure~\ref{fig:design_overview_researchers} depicts the workflows associated with these two evaluation tasks. 

\subsubsection{Data capturing.}
At the center of the \emph{data capturing} task lies the capturing devices.
We envision many capturing device types available in the \sys, which the AR researchers can choose from. 
For example, if the AR researcher is interested in capturing the ground truth for the lighting estimation task, she can select the capturing device that consists of $360^\circ$ cameras.
Note that the AR researcher does not need to be physically coupled with the capturing devices. 
Rather, the AR researcher will use any provided \sys client-side interfaces, e.g., a web UI, to control the movement of the capturing devices to mimic how an AR user will interact with the physical environment.

Moving the capturing device to different physical locations will allow the AR researcher to capture desired experimental data. 
These data, e.g., in the form of RGB video and movement commands, will be stored for later use. 
The AR researcher can start the data capturing \emph{in parallel} with different devices as the physical resource permits. 
To provide full control of the physical scene data, we will only allow at most one AR researcher to use the physical deployment. 
This resembles the current experimental practice where researchers are in charge of setting up the scene and introducing known dynamics to the scene.
In other words, we will impose the time sharing at the physical scene level, and the researcher can capture at most $\sum_i^m n_i$ stream of data in parallel where $m$ is the number of idle physical scenes, and $n_i$ is the number of capturing devices of $i^{th}$ scene. 
In addition to the parallel capturing, we will also allow researchers to monitor the capturing progress to spot any abnormalities, e.g., due to misconfigurations. 
In short, an AR researcher can use \sys to capture large amounts of experimental data from diverse scenes with heterogeneous devices to suit their evaluation needs.
Even better, the data capturing can be done with minimal human effort and does not require the costly acquisition of specialized devices. 

\subsubsection{Experiment design.}
\label{subsubsec:researcher_exp_design}

AR researchers can use \sys to setup the evaluation pipeline so that user study participants can use this pipeline to answer survey questions. Often the process involves AR researchers leveraging tools to create and generate visual assets for the survey questions and online survey platforms such as Quatrics to distribute the surveys. In other words, the survey question preparation and distribution are done in separate pipelines. This current practice works but can be time-consuming for AR researchers to design the survey questions. Instead, our goal in designing \sys is to streamline the experiment design process by allowing AR researchers to ``plug-and-play'' different pipeline components to assemble final visual products used in the survey questions. For example, AR researchers can take a raw video stream and pass it through an end-to-end pipeline to generate a post-processed video stream in which they can directly embed questions to suitable frame locations.

More concretely, with \sys, the researcher will come up with a list of questions, and use these questions to guide the configuration of the evaluation pipeline. For example, if the researcher is interested to evaluate the performance of her rendering-related AR tasks such as lighting and depth estimation, she can choose the \emph{baseline methods} hosted by \sys to render virtual assets, e.g., racing cars, in the same physical scene capture. With the rendered results, she can then create survey questions that ask participants to compare the relative rendering performance between her proposed method and a baseline method. Because the researcher is evaluating an AR task in isolation, we refer to this type of evaluation as \emph{standalone evaluation}. 

Researchers can also perform \emph{holistic evaluation} to study how well the proposed method works with the remaining sensing-perception-rendering pipeline. In holistic evaluations, \sys will ask researchers to select from predefined AR application scenarios such as museum tours or furniture shopping or configure their custom scenario. Each scenario specifies the AR tasks that need to be activated in the pipeline; for example, in the furniture shopping scenario~\cite{Yi2020-na}, \sys will activate the baseline models for image segmentation, object detection, and hand tracking tasks. For custom scenarios, the researchers are responsible for specifying interested AR tasks and their execution dependency. Afterward, researchers can leverage \sys to compare the end-to-end performance and quality trade-offs between the proposed method and any baselines. 

\sys will provide relevant data such as physical scene recordings (either directly recorded by this researcher or shared by others), 3D assets, and baselines for AR tasks. In addition, \sys will also allow researchers to upload any custom data and configure the data's visibility, private or public. Based on the evaluation questions, these data will be assembled \emph{on-demand}, providing the flexibility to conduct plug-and-play AR evaluations in scale.

\subsection{Design for AR User Study Participants}

This section describes the design of \sys by explaining how a user study participant will use \sys to complete two types of common evaluations: survey questions and participant observations.

\subsubsection{Participant observation.} It is valuable to understand how end users interact with a new AR system. However, it can be troublesome and time-consuming to invite user study participants to a physical experimentation site. Other extraordinary conditions such as the COVID-19 pandemic can also limit the practice of this experimentation form. For multi-user collaborative AR systems, it can be even more challenging to invite multiple AR users to be physically present in the experimentation site simultaneously. Furthermore, requiring physical presence also restricts the demographic diversity of the participants as people who are physically close by, e.g., university students, are more likely to participate in the study. 

\sys aims to facilitate the participant observation experiments by relaxing the physical presence requirement. That is, AR users can remotely interact with the AR systems (and devices). This concept is similar to RoboTurk, a recent robotic framework that allows human users to demonstrate how to perform tasks~\cite{roboturk}. Because of the relaxed physical presence requirement, \sys can then support more geographically diverse user study participants and make multi-user experiments easy to conduct. To allow AR researchers to observe the interactions, \sys will provide a web portal that in real-time displays AR users' interactions, an over-the-head view of the physical scene and the AR devices' movement, and the video streams from individual AR devices. Moreover, \sys will store those data in the backend to support post-analysis. 

In short, \sys will need to support streaming device perception from a physical scene to where the AR user locates and then send the AR user's interactions back to the physical devices. This communication pattern is similar to cloud gaming, and therefore we suspect the challenges lie in designing network optimizations to provide low-latency interactions.

\subsubsection{Online survey.} Understanding how a human user perceives AR features is important, and such understanding is often achieved via user studies. A common way to conduct user studies~\cite{Prakash2019-gb,Zhang2022-lq} is to invite participants to answer survey questions, e.g., how two competing techniques compare visually. One of the key features \sys can support is to allow user study participants to take these surveys via the web portal. As described in \S\ref{subsubsec:researcher_exp_design}, these survey questions are designed ahead of time by AR researchers who will then invite participants by sharing the survey URLs. Depending on the survey questions, the participants might be watching a replayed video stream in which frames are overlayed with information such as environmental conditions and visual outputs from different AR models. Relevant survey questions will be displayed to the participants at prespecified frames and responses will be collected, similar to how MOOC quizzes online learners' understanding of course topics.
\section{Reflection Questions}
\label{sec:reflection_qs}

\subsection{Group One}
\begin{itemize}[leftmargin=.12in,topsep=0pt]
    \item \emph{Describe the general process you took when designing and running the experiments in one of your recent AR works.}
    \item \emph{How long did setting up the physical testbed take?} 
    \item \emph{How long did data capturing take?}
    \item \emph{How long did obtaining results from baselines take?}    
    \item \emph{How many mobile devices did you evaluate your system on?}
    \item \emph{What were some reasons that prevent you from evaluating your system on more mobile devices?}
    \item \emph{What are the task(s) in evaluating the AR systems that you find yourself repeatedly doing all the time?}
    \item \emph{Was this experience described above, the same as other research projects you have done? The same as other AR research projects?}
\end{itemize}

\subsection{Group Two}

\begin{itemize}[leftmargin=.12in,topsep=0pt]
        \item \emph{If you could have access to a magical experiment setup that would allow you to do anything you want to evaluate your system, what would this experiment setup look like?}
        \item \emph{How would it work?} 
        \item \emph{What would you use this magical setup to do?}
\end{itemize}

\section{Module Implementation Details}
\label{sec:module_level_impl_details}

\subsection{Front end}

We implemented a web UI to support the two stakeholders of \sys. Currently, it consists of three components: 
\1 The \emph{Capture Viewer} component allows the AR researchers to watch a live stream from the capture device in the data collection process.
\2 The \emph{Session Rewatch} component fetches the processed frames from the cloud storage and presents them to the user for user studies, allowing assessment and analysis of previously recorded AR sessions.
\3 The \emph{Device Control} component processes controls from researchers, e.g., in the form of configuration files, and then controls various data capturing devices to initiate the data capture process.
In the future, we also plan to support other UIs, such as VR headsets and controllers.

\subsection{Capturing Device}

This module interfaces with the hardware sensors and the onboard computational resources. 
Currently, the hardware consists of a 360$^{\circ}$ camera, a PiCar-X, and an RPi 3B+. 
Our modular design allows for integrating additional capturing devices, such as drones and mobility-enabled specialized devices, in the future. 

We implemented two key modules: the \emph{FastAPI server} for processing user control commands and \emph{FFmpeg} for streaming video frames in MJPEG to the front/back ends. 
We have FFmpeg send the incremental frame counter to the FastAPI server to synchronize the time between controls and the video frames.
Specifically, the streaming functionality was implemented by having a process to read PiCamera frames directly from the camera and then send them to the backend through FFmpeg via an RTSP connection executed in a subprocess. We implemented a continuous frame counter that increments with each new frame captured and shared the counter across all tasks. 

We also used FastAPI to create a WebSocket connection between the front end and the capturing device. Once this connection is established, the live video feed will be activated and can be streamed to the back end. Moreover, the WebSocket connection remains open and actively listens for user inputs. These inputs, formatted as device control instructions, are forwarded to the device interface for execution.

\subsection{Back end}

Our back end consists of four main components for storing, processing, and streaming the captured data. 
The back end was implemented as Docker microservices, making the setup easily reproducible.
The back end processes the inbound data flow from the capturing device, including video, audio, and metadata.

The \emph{MediaMTX} server encodes the raw MJPEG stream into H.264, a format that can be streamed to the front end. It also generates the video segments for the \emph{data packing} component and interacts with the \emph{inference} service to augment the video stream.
The \emph{Redis} component logs user actions, which will be supplied to the \emph{data packing} component.

The \emph{Data Packing} component post-processes and consolidates all the data recorded during an AR session into a single MP4 file, at the end of each session. It runs a Redis pub/sub subscribe listener and will start a background process when messages are published to the data packing channel. It retrieves events from the Redis stream log and encodes them as JSON text to generate a SubRip Text (SRT) file. We use SRT, rather than KLV, to include synchronized metadata in the streams because KLV does not have good open-source support. It then uses FFmpeg to combine video segments and the SRT file into a single MP4 file. Inference output data can also be encoded as subtitles. 

The \emph{Inference} component runs DL models on the video stream. Currently, we implemented a popular object detection model called YOLOv8 as a proof-of-concept. Upon initiation by MediaMTX, it receives the video stream and applies the YOLOv8 model to each video frame. It returns the video to MediaMTX, which subsequently streams it back to the client and stores it for future playback purposes.
Our modularized design makes integrating other AR models into \sys as microservices easier.

\section{Resource Utilization Results}
\label{subsec:additional_results:resourceutil}

To better understand the performance bottlenecks of \sys, we measured the resource utilizations under different hardware setups. 
To facilitate these tests, we developed a mock car script that emulates car movements. This enables us to capture and stream content on the three computation devices while simultaneously measuring system resource usage. 

\begin{figure}[t]
  \centering
  \begin{subfigure}[b]{0.23\textwidth}
    \includegraphics[width=\textwidth]{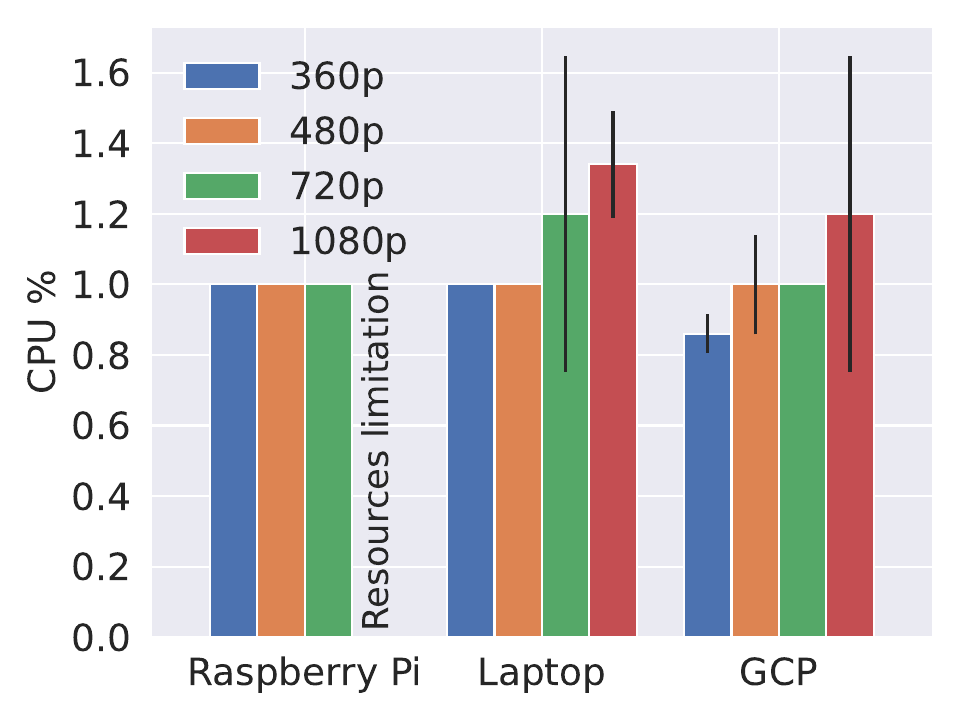}
    \caption{CPU Uvicorn}
    \label{fig:image1}
  \end{subfigure}
  \begin{subfigure}[b]{0.23\textwidth}
    \includegraphics[width=\textwidth]{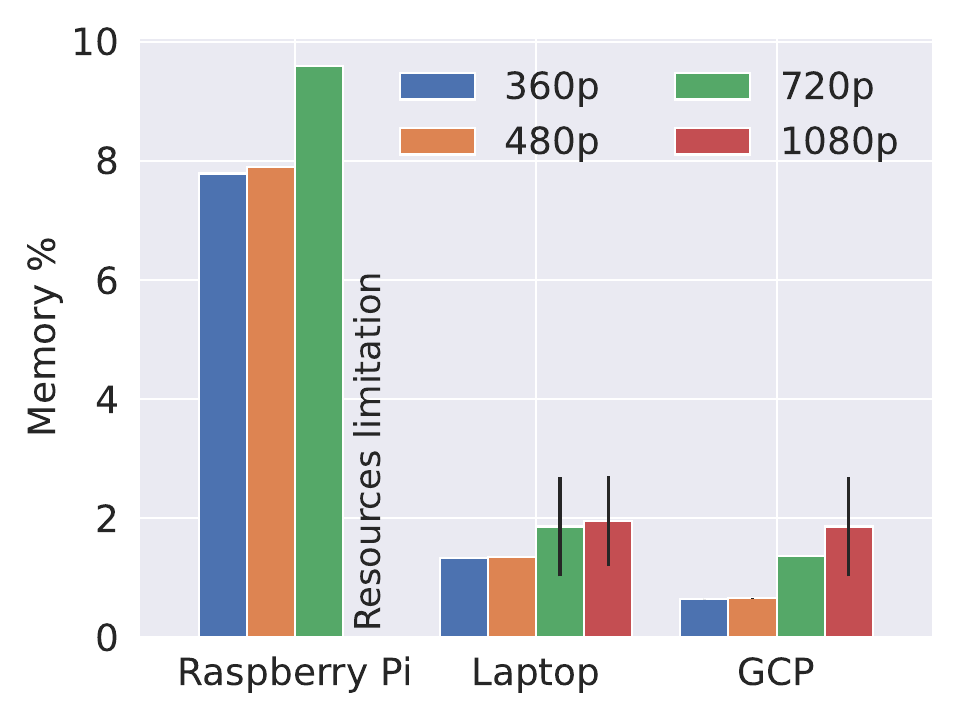}
    \caption{Memory Uvicorn}
    \label{fig:image2}
  \end{subfigure}
  \begin{subfigure}[b]{0.23\textwidth}
    \includegraphics[width=\textwidth]{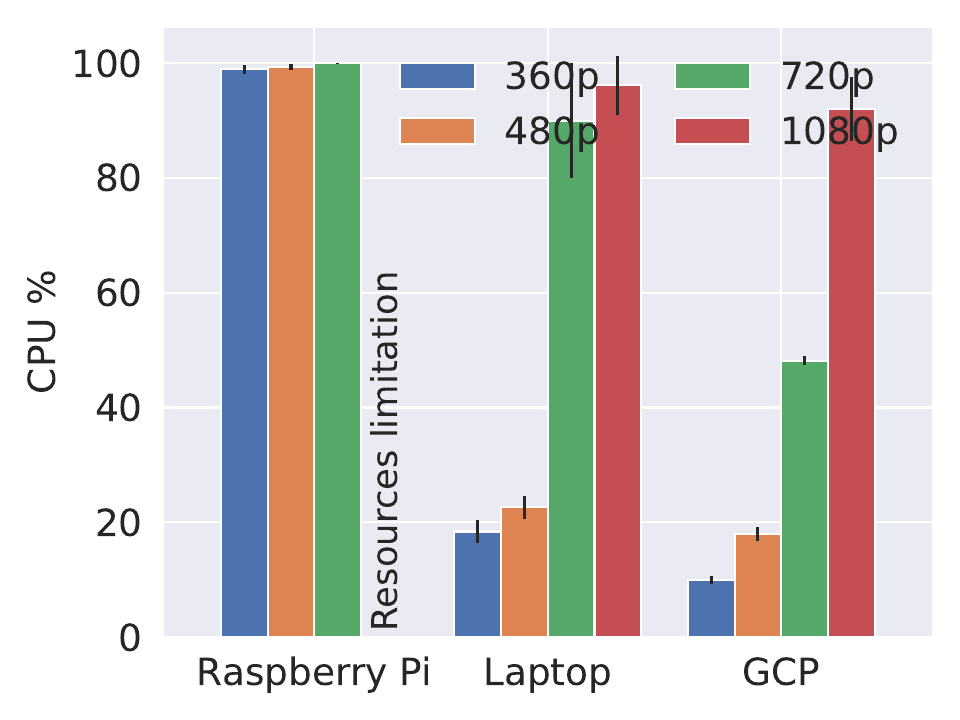}
    \caption{FFmpeg CPU}
    \label{fig:image3}
  \end{subfigure}
  \begin{subfigure}[b]{0.23\textwidth}
    \includegraphics[width=\textwidth]{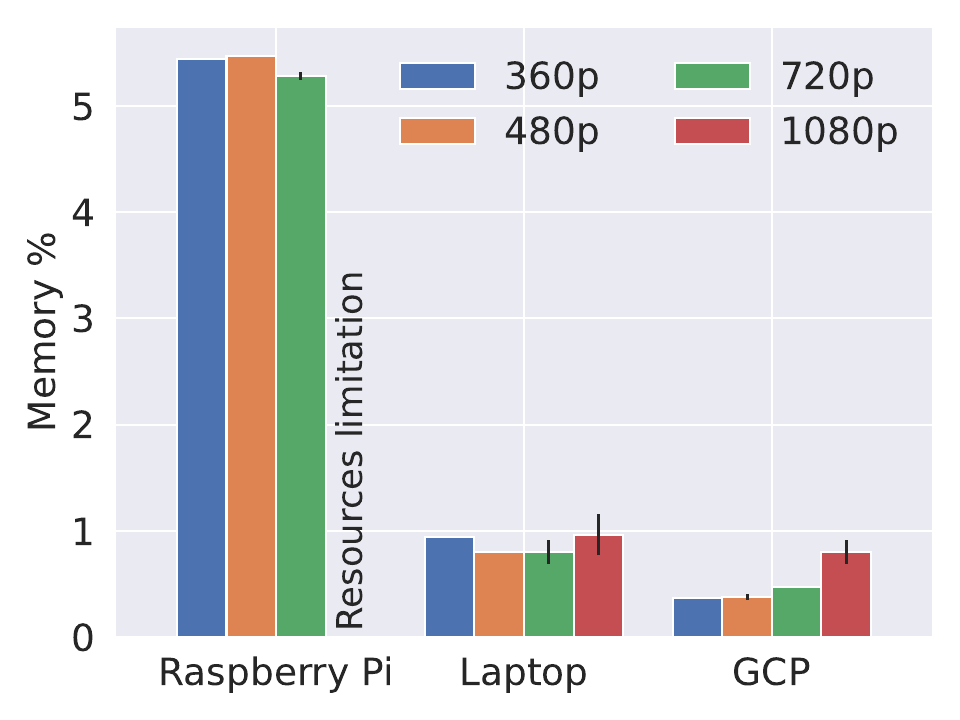}
    \caption{FFmpeg Memory}
    \label{fig:image4}
  \end{subfigure}
  \caption{
  Resource utilization comparison. 
  \textnormal{We measure both CPU and memory utilization when streaming using residential Wi-Fi. We see that FFmpeg consumes significantly higher CPU as resolution increases.
  }
  }
  \label{fig:system-resources}
\end{figure}

As shown in Figure~\ref{fig:system-resources}, higher video quality consumes more resources, especially by the FFmpeg process, which is responsible for stream and video format conversion. The Raspberry Pi struggled with high-quality video (1080p) processing, with FFmpeg maxing out CPU usage, and caused watchdog reset. Our results on the laptop and cloud server had a similar trend, with the CPU utilization increasing with the resolution. For higher resolutions, FFmpeg is again experiencing CPU bottlenecks.
This indicates that the current implementation of \sys is not effectively utilizing system resources.
Raspberry Pi's performance limitations and underutilization of resources in more powerful hardware highlight future improvement areas in exploring techniques to improve resource utilization and ensure smooth, high-quality AR experimentation experiences on a wide array of hardware.

\end{document}